\documentclass[
  a4paper,
  fontsize=11pt,
  ]{article}

\pdfoutput=1

\usepackage{pos_ILD}
\usepackage{graphicx}

\usepackage[utf8]{inputenx}
\usepackage[T1]{fontenc}

\usepackage[british]{babel}
\usepackage{csquotes}

\usepackage{subcaption}
\usepackage{import}
\usepackage{xspace}
\usepackage{graphicx}
\usepackage[shortcuts]{extdash}
\usepackage{siunitx}
\DeclareSIUnit \lightspeed {\text{{c}}}
\usepackage{xcolor}
\definecolor{linkblue}{HTML}{264772}
\usepackage{hyperref}
\hypersetup{
    colorlinks=true,
    linktocpage=true,
    linkcolor=linkblue,
    citecolor=linkblue,
    urlcolor=linkblue
}
\usepackage[noabbrev,capitalise]{cleveref}

\usepackage{wrapfig}

\graphicspath{ {../pictures/} }
\DeclareGraphicsExtensions{.pdf,.png,.jpg}

\newcommand{\dEdx}{\ensuremath{\mathrm{d}E/\mathrm{d}x}\xspace}

\title{The International Large Detector (ILD) for a future electron-positron collider: Status and Plans}

\manuallySeparateAuthors
\author*[a]{Ulrich Einhaus}
\author{ for the ILD Concept Group}

\affiliation[a]{Deutsches Elektronen-Synchrotron DESY, Notkestr. 85, 22607 Hamburg, Germany}

\emailAdd{ulrich.einhaus@desy.de}

\abstract{
This work presents the status and plans of the International Large Detector (ILD) concept, one of the most detailed and comprehensive detector concepts for a future Higgs factory.
Most hardware groups have demonstrated ILD's performance targets and continue development with  focus on improving further and making ILD fit for a circular collider.
Their status, new developments and plans are elaborated.
Two examples are given of new reconstruction methods that utilise hardware developments and contribute to advanced physics analyses prospects.
}

\FullConference{The European Physical Society Conference on High Energy Physics (EPS-HEP2023)\\
 21-25 August 2023\\
Hamburg, Germany\\}

\begin{document}
\maketitle

\newpage
\section{Introduction}

The area of future particle accelerators today is wider than it has been for many years.
While the particle physics community agrees that the next collider should be an e$^+$e$^-$ Higgs factory, a number of proposals for accelerators as well as detectors have been brought forward.
The International Large Detector (ILD) concept, developed for the International Linear Collider (ILC), is one of the most advanced concepts for a detector at a future Higgs factory.
For more than 15 years now, ILD has been a driver of hardware, software and physics analysis developments in the field.
This work gives an overview of the status and future plans of ILD.
It highlights recent developments in hardware as well as reconstruction software and their combined impact on physics benchmarks.

ILD was developed as a multi-purpose detector for ILC's \cite{ILC_TDR_short} center of mass energies of 250 GeV, 500 GeV and 1 TeV.
It is optimised for ParticleFlow \cite{Pandora}, including using a time projection chamber (TPC) as main tracker, which allows for continuous tracking as well as particle identification (PID) via \dEdx and minimises the material budget in front of the calorimeter system, that is placed entirely inside the 3.5 T solenoid. This is complemented by a near-4$\pi$ hermeticity with an elaborate forward tracking and calorimetry system.
The most recent comprehensive overview of the detector concept is the Interim Design Report \cite{ILD_IDR}, including hardware status as well as detector performance and physics prospects.
These prospects are based on large MC productions \cite{Production_2020} using full Geant4-based simulation and reconstruction. The corresponding software has become a central contribution to the Key4hep framework \cite{Key4hep} for future Higgs factories. The most prominent detector performance figures and comparisons to currently operating LHC detectors leading in that aspect are given below.
Recently, ILD started studying if it is feasible to be proposed for the FCC-ee, where operating conditions are very different: continuous beams instead of trains, less beamstrahlung, final focusing magnets closer to the IP and a lower magnetic field.

\begin{table}[h!]
 \begin{tabular}{llr}
  Vertexing: & $\sigma_b < (5\oplus10/p \sin^{3/2}\theta)\mu m$ & $\sim$ CMS/4 \\
  Momentum: & $\sigma_{1/p_{T}} = 2\cdot 10^{-5} GeV^{-1}\oplus 10^{-3}\sin^{1/2}\theta/p$ & $\sim$ CMS/40 \\
  Jet energy: & $\sigma_{E_{Jet}} / E_{Jet} < 3.5\% $ \hspace{.4cm} above 100 GeV & $\sim$ ATLAS/2 \\
  \dEdx : & $\sigma_{dE/dx} / \mu_{dE/dx} < 5\% $ & $\sim$ ALICE\\
  
 \end{tabular}
\end{table}

\begin{figure}[!hbt]
  \centering
    \includegraphics[width=.9\textwidth,keepaspectratio=true]{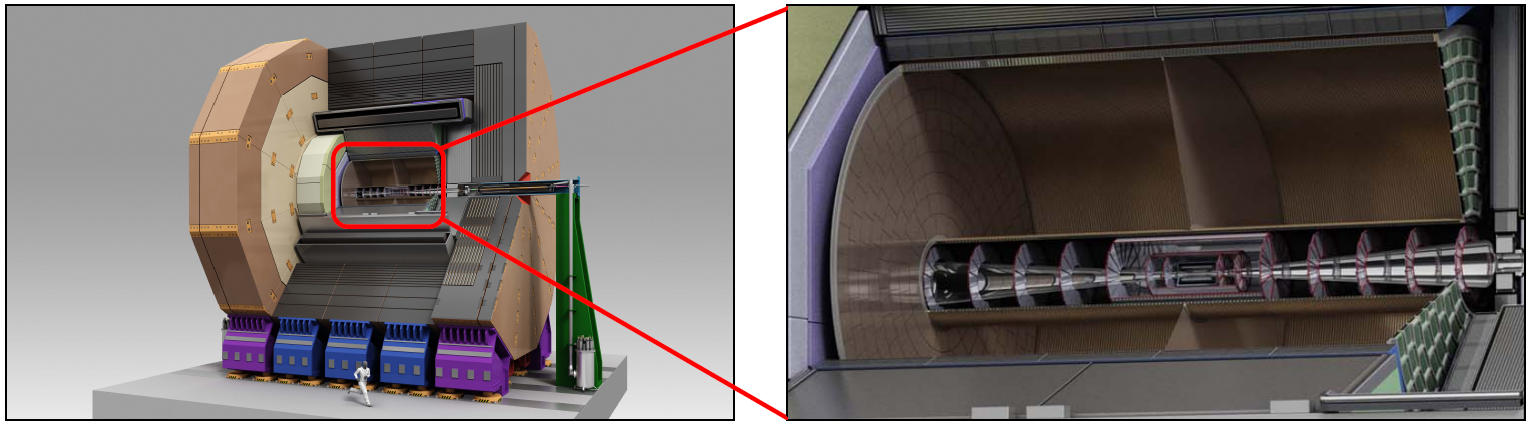}
    \caption{Artist render of ILD \cite{ILD_IDR}.}
  \label{fig:ILD}
\end{figure}

\vspace{-.5cm}

\section{Hardware R\&D}

While ILD foresees the `usual' approach to a Silicon vertex detector, this is a very active field and it will be one of the last sub-detectors for which a design needs to be fixed, hence a number of options are under active consideration, such as CMOS, DEPFET, MAPS and (i)LGADs.
A recent development of particular interest is the ALICE ITS3 \cite{ITS3}, that uses bendable MAPS and has only 0.05\% of a radiation length per layer while providing a point resolution of better than \SI{5}{\mu m}. This performance of an existing vertex detector highlights the possibility of updating the current ILD vertexing performance prospects, which appear increasingly conservative.
Another notable development are LGADs and their derivatives such as inverted LGADs (iLGADs) \cite{iLGADs}.
These sensors provide a timing resolution of a few \SI{10}{ps} and the ILD detector model has been updated to include a timing measurement in the Silicon envelope of the TPC or the first ECal layers, which would enable a time-of-flight (TOF) measurement.
The name-giving inversion of iLGADs is more complex since both sides of the wafer need to be processed, but allows to maximise the fill factor by segmenting the hole-collecting side instead of the amplification side.
\\

The development of the TPC, in particular its readout, is driven by the LCTPC collaboration \cite{LCTPC_2023}.
Readout systems based on gas electron multipliers (GEMs) and MicroMegas have been developed and demonstrated, achieving the performance goals of ILD.
While these systems use mm-sized readout pads, newer developments concentrate on high granularity, utilising the 55x55 $\mu$m$^2$ pixels of the Timepix ASIC as readout anode.
Large modules of this GridPix system for a `PixelTPC' have been developed and demonstrated.
Extrapolations to ILD show that the momentum resolution could be improved by about 15\%, the effective \dEdx resolution by about 30\% using the cluster counting method \cite{JointPaper}.

Proposing ILD for the FCC-ee puts in the TPC somewhat in question.
Continuous running does not allow for active gating and high rates both from hadronic physics at the Z pole as well as from beamstrahlung due to the material of the final focusing quadrupole being closer to the IP introduce a large amount of slow ions in the TPC sensitive volume.
This space charge distorts the drifting electrons ad worsens the momentum resolution.
To study this, an ILD detector model has been devised that contains the forward region of the FCC-ee detectors \cite{TPC_CC}.
The amount of ions per bunch crossing is similar between ILC and FCC-ee conditions, since the additional material in the FCC-ee forward region is compensated by the reduced amount of beamstrahlung due to less focused beams.
However, the higher bunch crossing frequency at a circular collider lets the ions accumulate.
With FCC-ee conditions at \SI{240}{GeV}, the amount of ions would be about 200 times of ILC at \SI{250}{GeV}, at FCC-ee \SI{91}{GeV} about 2500 times.
The latter amount would cause distortion of a few cm.
Work is ongoing to assess how stable these distortions are, i.e.\ how well they can be corrected for, how much of the ions can be mitigated by adapting the FCC-ee forward system and how much less strict the performance requirements are at the Z pole compared to Higgs production.\\


For the main calorimeter systems the CALICE collaboration \cite{CALICE_Overview} has developed several options which all target high granularity, i.e.\ 5x5 mm$^2$ cells in the ECal and 10x10-30x30 mm$^2$ cells in the HCal, which means about $10^8$ readout channels in total.
Due to the bunch train structure of the ILC, the electronics' bias currents of many ILD sub-systems can be shut down during idle times to save power, so-called power pulsing. This allows to run the calorimeter systems without active cooling, which assures the compactness and hermiticity of the detector.
For the ECal there are the options of using tungsten absorber and Silicon sensors (SiW) with large surface silicon wafers (90x90 mm$^2$) segmented in 5x5 mm$^2$ individual pads, while the scintillator (Sci) approach uses 5x45 mm$^2$ strips, which provide the granularity by stereo angle and would reduce the channel count by a factor 10.
The analogue HCal (AHCAL) uses scintillator and SiPMs, while the semi-digital HCal (SDHCAL) uses RPCs.
For all systems prototypes exist and have been beam tested.

Recently, SiW-ECal and AHCAL have conducted a common test beam campaign that used a common slow control and readout. These two CALICE-developed systems have also inspired the HGCAL which is set to be installed as CMS endcap calorimeter for the HL-LHC.
In addition, the SiW-ECal group is working on a new PCB fully optimised for power pulsing to finish the ILD-ready design.
The group is also looking into a possible application of the SiW-ECal at the LUXE experiment.
The Sci-ECal group is working towards a common test beam with the HCal of the CEPC detector.
AHCAL works on potentially using tungsten absorber instead of steel and continues the development of so-called Megatiles, which would simplify the production process.
The SDHCAL is looking to also build an ILD-sized (1x1x3 m$^3$) prototype and is part of the CEPC HCal group.
A major topic that all group are working on is the ability to cope with continuous operation at a circular collider, which affects bandwidth, power consumption and cooling.
In addition, timing capabilities will be integrated into all calorimeter systems, which will allow for TOF measurements as well as 5D ParticleFlow.
\\

The detailed forward-system of ILD has been developed by the FCAL collaboration \cite{FCAL}.
It entails the LumiCal, which provides a precise luminosity measurement by counting Bhabha events, and the BeamCal, which conversely provides a fast luminosity measurement using beamstrahlung.
In combination, the FCal system covers the polar angle down to 6 mrad ($\eta$ = 5.8) and is the cornerstone of ILD's hermeticity, which is important not only for ParticleFlow but also for new particle searches of invisibles via pick up of ISR.
A design with detailed simulations has been done and a prototype of the LumiCal with Silicon sensors has been built and tested, with more prototypes envisaged.
While spin-offs of the FCal went into the CMS luminometer and beam condition monitor, a possible future application is LUXE.
A recent development is GaAs sensors with integrated routing.

\section{Reconstruction Algorithm R\&D for Physics}

Another part of research for future colliders is the development of new reconstruction algorithms, their interconnection and the study of their detector dependence on the one hand and their impact on the physics performance on the other hand.
Two such developments for ILD are highlighted in the following.
\\

\begin{wrapfigure}[14]{R}{.48\textwidth}
  \centering
    \vspace{-.5cm}
    \includegraphics[width=.48\textwidth,keepaspectratio=true]{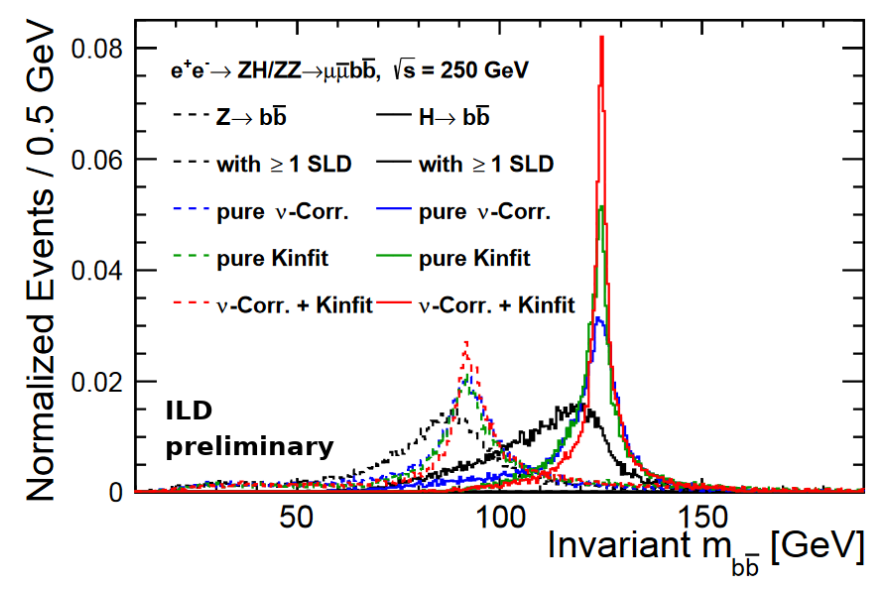}
    \caption{Reconstruction of invariant Z and Higgs mass utilising neutrino correction and ErrorFlow, from \cite{ICHEP_2022}.}
  \label{fig:HZ_SLDs}
\end{wrapfigure}
One important aspect of measuring Higgs properties is to reduce background as much as possible, in particular from the Z peak.
This Z/H separation depends on the resolution of the invariant masses of the corresponding bosons.
This is significantly affected by neutrinos in the final state which carry away invisible energy.
These occur in particular in semi-leptonic decays (SLDs) of b and c hadron in Higgs or Z decays to $b\bar{b}$ - the single largest branching fraction of the Higgs - and $c\bar{c}$.
This is of particular interest for the Higgs self-coupling measurement and the separation of ZHH from ZZH events \cite{Torndal_2023}.
\autoref{fig:HZ_SLDs} \cite{ICHEP_2022} illustrates the impact of SLDs for ZZ/ZH to $\mu\bar{\mu}b\bar{b}$ events in the black lines for events with at least one SLD, where Higgs and Z peak are smeared out and shifted to lower masses due to the missing energy.
Given the known initial state of the collision, the missing neutrino momentum can be compensated, which has been demonstrated in a new study \cite{Yasser_2023}.
For the calculation, it is also necessary to find all outgoing charged particles, correctly associate them to the corresponding b- and/or c-hadron vertex and successfully flavour tag these vertices.
This requires high detector hermeticy and excellent vertexing.
Then the neutrino momentum can be fully reconstructed - up to a final sign ambiguity.
With a cheated sign, one receives the blue curves, which are already much narrower and have a correct peak position.
In addition, a kinematic fit can utilise certain event constraints to re-assess the measured jets and their constituents.
While this method has already been used at LEP, a new approach, called ErrorFlow, uses individual uncertainties of each reconstructed ParticleFlow object (PFO) instead of generic ones to improve on the re-assessment.
After combining the uncertainties of the individual PFOs into a covariance matrix for the reconstructed jets, this is performed on jet level.
ErrorFlow alone leads to the green curves, and a combination of neutrino correction and ErrorFlow, which then also takes care of the sign ambiguity, to the red curves.
Here, the Z peak is dominated by the intrinsic width of the Z and the remaining background for the Higgs - based on invariant mass - has been minimised.\\

In recent years, PID has sparked increasing interest.
In particular the developments of O(10) ps timing in Silicon for the LHC experiments opened the possibility for PID via TOF.
This has been studied and fully implemented for the detailed ILD detector simulation \cite{TOF_EPS23}.
This new implementation uses reconstructed momentum and track length utilising a hit-by-hit calculation inside the Kalman filter of the track fit in addition to a timing measurement with a variable resolution.
This allows for a hadron mass reconstruction of unprecedented realism at future collider detector concepts.
One conclusion from these studies is that with assumed 30 ps-timing the track length has a similar impact on the mass reconstruction as the timing itself.
Since this was performed with the up to 220 continuous hits per reconstructed track of the ILD TPC, it raises the question what track length precision full-Silicon trackers can achieve and if this satisfies requirements for TOF.
This is even more relevant, since in full-Si detectors TOF may be the only viable PID option.

In ILD, however, the TPC also provides \dEdx PID, which is currently combined with cluster shape information from Pandora to give best estimates for charged particle ID.
To allow for an easy addition, combination and comparison of different PID observables, a new framework, called Comprehensive PID (CPID), was recently developed.
It uses a modular approach both in the PID observables as well as in the training models that are used to combine the different observables.
The inclusion in Key4hep now allows to not only optimise individual detectors and their subsystems but also to compare different proposed detector concepts in a coherent way.
While numerous studies have shown the impact of PID on the physics case of future Higgs factories, one new study \cite{AFB_2023} highlights in particular ILD TPC.
A measurement of the hadronic s-channel ($e^-e^+ \rightarrow q\bar{q}$) forward-backward asymmetry $A_{FB}^q$ for different quark flavours as well as the corresponding hadronic Z branching fractions $R_q$ is sensitive to a number of BSM models.
In b and c jets high-momentum charged kaons carry crucial information if the initial quark was a particle or antiparticle, a central question for the forward-backward asymmetry. 
The study shows the impact of the TPC's \dEdx on this measurement as well as on the possibility to finally distinguish various BSM models not only from the Standard Model but also from each other.

\begin{figure}[!h]
  \centering
    \includegraphics[width=.5\textwidth,keepaspectratio=true]{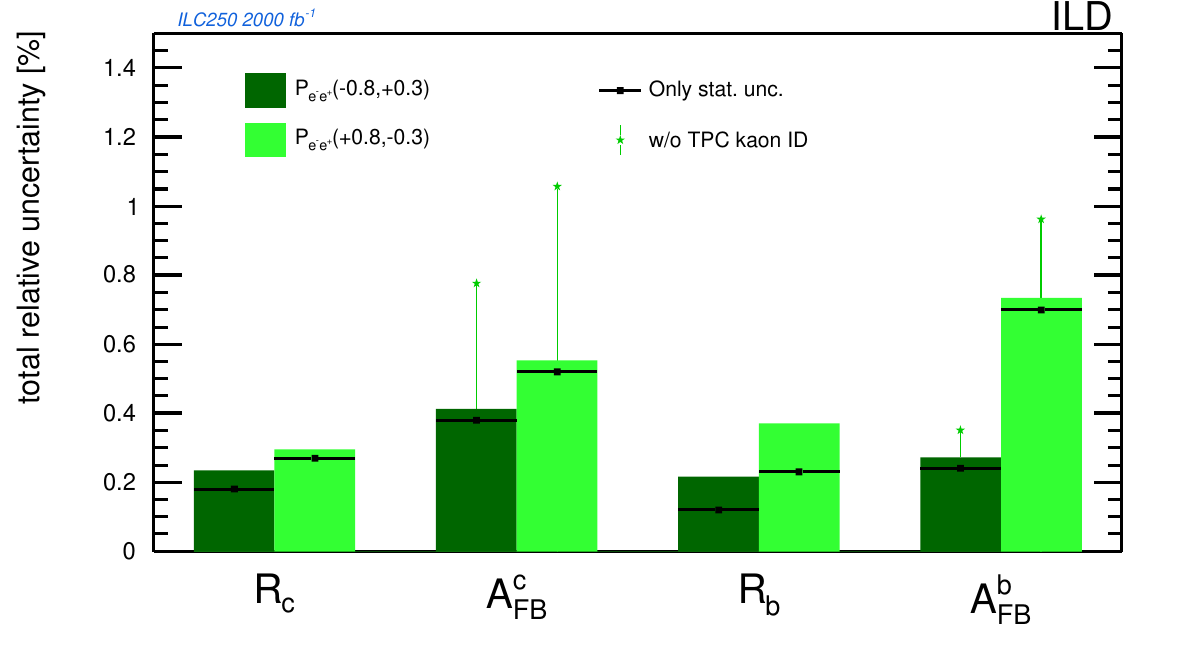}
    \hspace{1cm}
    \includegraphics[width=.27\textwidth,keepaspectratio=true]{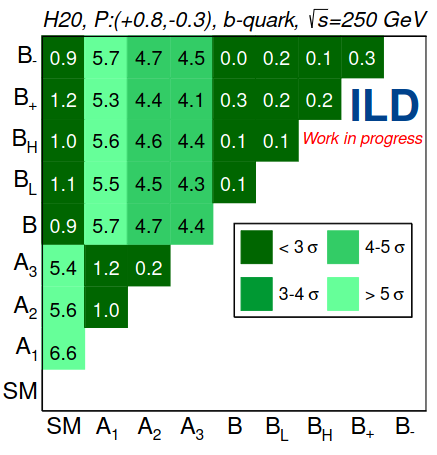}
    \caption{Left: Impact of kaon ID with TPC \dEdx on the $A_{FB}^q$ measurement, from \cite{AFB_2023}; right: sensitivity of this measurement to BSM models, from \cite{AFB_2023b}.}
  \label{fig:AFB}
\end{figure}

\vspace{-.5cm}

\section{Conclusion and Outlook}

Thanks to its age and its still continued activity in developing hardware, software and physics prospects, ILD is one of the best-understood detector concepts for a future Higgs factory.
Its performance targets have largely been demonstrated by large prototypes, but development continues and focuses now on improvements beyond what was considered ambitious 15 years ago as well as adaptations to beam conditions at a circular collider.
Hardware developments inform physics opportunities for which new reconstruction algorithms are needed, such as highly granular calorimeters and hermeticity for ParticleFlow that enables neutrino correction for Higgs precision measurements, or fast timing that enables TOF PID for BSM searches.
Through their integration in Key4hep the software developments of the linear collider community including ILD of many years with robust reconstruction algorithms are available to the broader future Higgs factory community.
ILD continues to contribute to this framework, with new developments often being addressed to a wider audience than a single detector.
Physics studies are ongoing and proceed to make the case for a future $e^+e^-$ machine.
ILD continues to be a driver in hardware, software and physics for a future Higgs factory.

\section{Acknowledgments}
I thankfully acknowledge the support by the Deutsche Forschungsgemeinschaft (DFG, German Research Foundation) under Germany’s Excellence Strategy EXC 2121 "Quantum Universe" 390833306.

\bibliographystyle{JHEP}
\bibliography{References_reduced}
\end{document}